\documentclass[%
 reprint,
superscriptaddress,
 amsmath,amssymb,
 aps,prl
]{revtex4-2}
\usepackage{graphicx,epsfig}
\usepackage{siunitx}
\DeclareSIUnit\molar{M}
\DeclareSIUnit\electronvolt{eV}
\usepackage{braket}
\usepackage{bm}
\usepackage{dcolumn}
\usepackage{xcolor}
\usepackage{color,soul}
\usepackage[breaklinks=true,colorlinks,citecolor=blue,linkcolor=blue,urlcolor=blue]{hyperref}
\usepackage[utf8]{inputenc}
\usepackage[T1]{fontenc}
\newcommand\ntwo{$N{-}2$} 
\newcommand\nfour{$N{-}4$} 
\newcommand\nthree{$N{-}3$} 
\newcommand\nfive{$N{-}5$} 

\begin{document}
\title{Charge-state dependent vibrational relaxation in a single-molecule junction}
\author{Xinya Bian}\affiliation{
Department of Materials, University of Oxford, Oxford, OX1 3PH, UK
}
\author{Zhixin Chen}\affiliation{
Department of Materials, University of Oxford, Oxford, OX1 3PH, UK
}
\author{Jakub K. Sowa}\affiliation{
Department of Chemistry, Rice University, Houston, TX, USA
}
\author{Charalambos Evangeli}\affiliation{
Department of Materials, University of Oxford, Oxford, OX1 3PH, UK
}
\author{Bart Limburg}\affiliation{
Department of Chemistry, University of Oxford, Oxford, OX1 3TA, UK
}
\author{Jacob L. Swett}\affiliation{
Department of Materials, University of Oxford, Oxford, OX1 3PH, UK
}
\author{Jonathan Baugh}\affiliation{
Institute for Quantum Computing, University of Waterloo, Waterloo, ON N2L 3G1, Canada
}
\author{G. Andrew D. Briggs}\affiliation{
Department of Materials, University of Oxford, Oxford, OX1 3PH, UK
}
\author{Harry L. Anderson}\affiliation{
Department of Chemistry, University of Oxford, Oxford, OX1 3TA, UK
}
\author{Jan A. Mol}\affiliation{
School of Physical and Chemical Sciences, Queen Mary University, London, E1 4NS, UK
}
\author{James O. Thomas}\email[Electronic address: ]{
james.thomas@materials.ox.ac.uk
}
\affiliation{
Department of Materials, University of Oxford, Oxford, OX1 3PH, UK
}

\begin{abstract}
The interplay between nuclear and electronic degrees of freedom strongly influences molecular charge transport. Herein, we report on transport through a porphyrin dimer molecule, weakly coupled to graphene electrodes, that displays sequential tunneling within the Coulomb-blockade regime. The sequential transport is initiated by current-induced phonon absorption and proceeds by rapid sequential transport via a non-equilibrium vibrational distribution. We demonstrate this is possible only when the vibrational dissipation is slow relative to sequential tunneling rates, and obtain a lower bound for the vibrational relaxation time of 8 ns, a value that is dependent on the molecular charge state.
\end{abstract}
\maketitle
Charge transport through nanostructures, and molecules in particular, is strongly influenced by coupling between electronic and mechanical degrees of freedom\citep{Mitra2004,Park2000,Sapmaz2006,hartle11,Burzuri2014,Lau2015}. Geometric differences between molecular charge states mean that electron transfer steps are accompanied by vibrational transitions, an effect encompassed by the Franck-Condon principle\citep{Koch2005}. Typically vibrational transitions are identified as steps in current that are equally spaced in voltage in current-voltage (\textit{IV}) measurements\citep{Burzuri2014,Lau2015,Sapmaz2006,Leturcq2009,Thomas2019}. Transport behavior that results from coupling to equilibrated vibrations has been extensively studied and well understood. For example, electron-phonon coupling strength and vibrational frequency are precisely extracted from the progression of the current step heights, and low-bias suppression in the strongly coupled condition due to negligible vibrational wavefunction overlap, known as Franck-Condon blockade, has been observed\citep{Burzuri2014, Lau2015, Leturcq2009}. Conversely, the effects of non-equilibrium vibrational distributions are less well-established. Interesting phenomena such as current-induced molecular heating and dissociation, giant Fano factors, and avalanche tunneling dynamics have been predicted theoretically\citep{Koch2005,Koch2005a,Leijnse2008,Lueffe2008,ke21,chen03} and subsequently observed, primarily in carbon nanotubes devices\citep{ward11,Huettel2009,Leturcq2009,Sapmaz2006,Lau2015}. 

Vibrational effects in single-molecule junctions are particularly prominent in the weak molecule-electrode coupling limit, as strong electron-electron interactions lead to Coulomb-blockade (CB) behavior and charge states are well defined. Transport through the molecule occurs either via resonant sequential tunneling outside the CB regime or via off-resonant cotunneling inside the CB regime. Like sequential tunneling processes, cotunneling processes can also excite vibrational states above a bias threshold through inelastic cotunneling\cite{galperin07}. If cotunneling-excited vibrational states can promote sequential tunneling processes, there is a loss of definition of the boundary between resonant and off-resonant tunneling regimes\citep{Lueffe2008, gaudenzi17}. Depending on the interplay between vibrational relaxation, cotunneling, and sequential tunneling rates, cotunneling-assisted sequential tunneling processes can lead to distinctive features in the transport spectrum, known as absorption sidebands, extending into the CB regime\citep{Koch2005}.

In this Letter we report the first experimental observation of such sidebands in a single-molecule device and confirm that they originate from a current-driven, non-equilibrium vibrational distribution. We calculate a lower threshold for the molecular vibrational relaxation time of 8 ns, and demonstrate that there is a connection between dissipation rate and molecular charge state, and by extension, its geometry.

\begin{figure}
\includegraphics{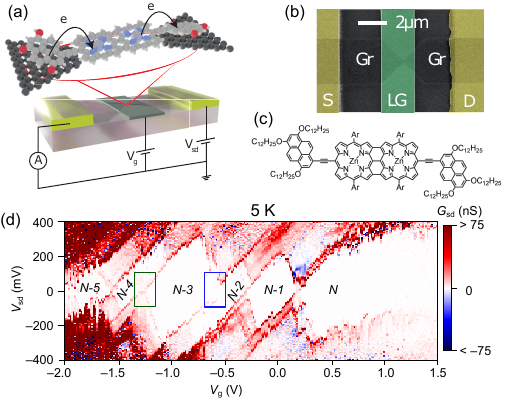}
\caption{(a) Device structure schematic. (b) False-color SEM image of graphene (Gr) transferred onto source (S) and (D) leads, and a local gate electrode (LG). Gr is patterned into a bowtie shape to produce nanoelectrodes via electroburning. (c) Molecular structure of \textbf{FP2}, Ar: 3,5-\textit{bis}(trihexylsilyl)phenyl. (d) Wide-range differential conductance map ($dI_{\rm{sd}}/dV_{\rm{sd}}$) of the \textbf{FP2} device, transitions studied in detail are outlined in green and blue.
\label{schematics}}
\end{figure}

The device architecture is displayed in Fig. \ref{schematics}, and has been described previously\citep{Lau2014}. A high-$\kappa$, $\SI{10}{\nano\meter}$-thick, HfO\textsubscript{2} gate dielectric gives a large electrostatic coupling between the molecular states and the gate potential, $V_{\rm{g}}$ (in this case $\alpha_{g} = 0.5$ eV/V). The molecule under study, a fused zinc-porphyrin dimer (\textbf{FP2}, Fig. \ref{schematics}c), has a small charging energy which, combined with the large gate coupling, allows multiple molecular charge states to be studied. \textbf{FP2} is synthesized by Sonogashira coupling of a dibromo- edge-fused zinc porphyrin dimer\citep{richert17} with an ethynylpyrene derivative\citep{limburg18}\footnote{See Supplemental Material for synthetic details, additional transport data, and fitting results.}. Pyrene groups anchor the molecule to graphene source and drain electrodes through a $\pi$-$\pi$ stacking interaction and provide weak molecule-electrode coupling, $\Gamma$. Electroburnt graphene nanoelectrodes can display transport features prior to molecular deposition and therefore devices were measured before and after deposition to ensure we only study transport features related to \textbf{FP2}.\footnotemark[\value{footnote}]. From 98 devices measured at 5 K, the majority displayed resonant tunneling features prior to molecule deposition, in line with previous work\citep{limburg18}. Five displayed clear resonant transport only after deposition and had consistent addition energies with the device studied in detail here. The presented device has the specific combination of molecule-electrode and gate coupling to allow the experimental study of absorption sidebands.

A large $V_{\rm{g}}$-range differential conductance map (Fig. \ref{schematics}d) confirms the molecule is weakly coupled to the graphene electrodes. Off-resonant transport is suppressed due to CB and multiple Coulomb diamonds, with addition energies in the range of 150--300 meV, are observed and separated by resonant transport regions. Due to the electron-rich anchor groups and \textit{p}-doping of graphene by the substrate\citep{ryu10}, charge states are assigned to successive oxidized states of \textbf{FP2}, i.e. \nfive~ to $N$, $N$ being the number of electrons on the molecule when it is neutral. The assignment and addition energies are inline with studies on the same family of molecules\citep{Thomas2019, limburg19, thomas21}. We focus on electron-vibration coupling features manifested in the charge transitions at $V_{\rm{g}} = -1.25$~V, assigned to the \nfour/\nthree~transition (green box), and $V_{\rm{g}} = -0.6$~V - \nthree/\ntwo~ (blue box).

\begin{figure*}
\includegraphics{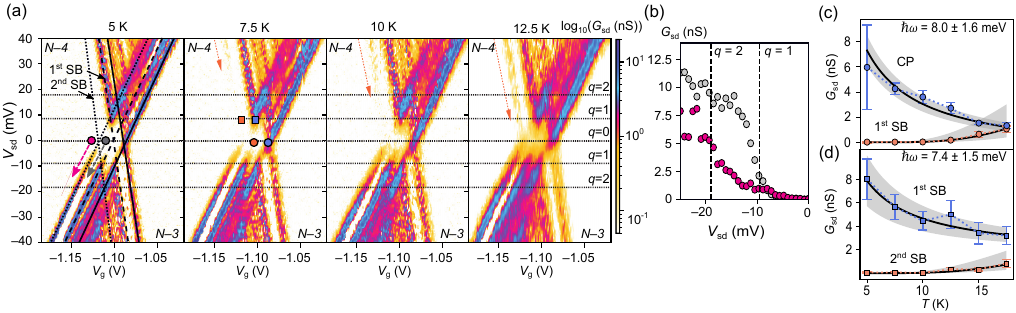}
\caption{(a) Differential conductance maps of the \nfour/\nthree~transition measured at different temperatures, showing quasi-periodic resonant transport features. The onset of the second sideband is indicated by the orange arrow. (b) Conductance measured along the first (grey marker/arrow on (a)) and second (pink marker/arrow on (a)) sidebands shows suppression below $eV_{\rm{sd}} < n\hbar\omega_{q}$. (c) Temperature-dependence of the Coulomb peak (CP) and the first sideband (SB), (red/blue circular markers on (a)). The ground state shows a simple $1/T$ temperature dependence (black line), the first sideband is suppressed, increasing due to thermal population of $q=1$, as fitted by \ref{eqn:bose} (black line). (d) Temperature-dependent conductance through the first and second sideband (square marker positions on (a)) at $eV_{\rm{sd}}=\hbar\omega_{q}$, the conductance of the first sideband is high and decays as $1/T$ (black line) whilst the second grows with temperature.} 
\label{temp_depen}
\end{figure*}

Fig. \ref{temp_depen}a displays high-resolution conductance maps of the \nfour/\nthree~transition. The solid slanted lines define the usual resonant transport region separating the \nfour~and \nthree~diamonds. Lines of increased conductance running parallel to the diamond edges are caused by additional transport channels to the ground-state to ground-state transition between \nfour~ and \nthree. The low energy of these excited-state channels, at around 9 meV (dashed/dotted lines), point to a vibrational origin. Vibrational transitions of a similar energy have been observed in a study of a porphyrin monomer\citep{Thomas2019}. Under weak coupling, resonant transport is dominated by first-order sequential tunneling processes, i.e. $\ket{N, q}$ to $\ket{N', q'}$ where $N' = N \pm 1$, and each additional channel involves either the hopping on/off steps required to transport an electron through the junction to be accompanied by a change in vibrational state, i.e. $\Delta q = q' - q \neq 0$. These steps generate out-of-equilbrium vibrational excitations in the molecule, but if the rate of vibrational relaxation is much faster than sequential electron transfer rates then under steady-state conditions the system can be treated as in equilibrium. Under equilibrium conditions and at low temperature, only the vibrational ground state is appreciably populated, CB is maintained, and sequential tunneling current should be strictly bound within the solid lines defining the ground-state to ground-state resonant tunneling transition. Here, however, more complex behavior is observed, and the additional transport channels at energies of $\hbar\omega_{q} = 9$~meV (the dashed/dotted lines slanted lines) do not terminate at the boundaries of CB regions but instead extend into the \nfour~ diamond, forming sidebands to the Coulomb peak. The \nfour~ sidebands run parallel to diamond edges, indicating current within them results from sequential tunneling processes. The first sideband corresponds to a transition with $\Delta q = -1$ and thus sequential tunneling must involve at least the first vibrationally excited state of \nfour. For the second sideband, $\Delta q = -2$ and sequential tunneling must originate from at least the second vibrationally excited state of \nfour. The sidebands are present at 5 K, despite the absence of thermal population in $\ket{N{-}4, 1}$. 

\begin{figure} 
\includegraphics{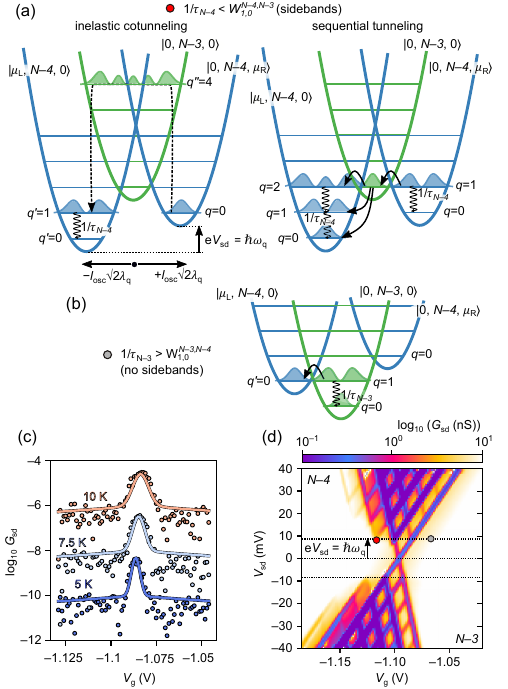}
\caption{(a) The onset of first sideband (red dot in panel (d)). Blue \nfour~ parabolas correspond to an electron at the Fermi level of left $\ket{\mu_{L}, N{-}4, 0}$ or right $\ket{0, N{-}4, \mu_{R}}$ electrode. The green parabola corresponds to the electron being on the molecule: $\ket{0,N{-}3,0}$. \nfour~parabola have the same $x$-position, but are offset from the \nthree~ by $\pm l_{osc}\sqrt{2}\lambda_{q}$ to visualise electron flow from right to left lead. Inelastic cotunneling promotes the molecule to $q=1$. If relaxation ($1/\tau_{N{-}4}$) is smaller than $W^{N{-}4,N{-}3}_{1,0}$, sequential tunneling proceeds via excited states of \textbf{FP2}, leading to sidebands. (b) Parabolas at the potential onset of the first \nthree~ sideband (grey dot) are not observed as $1/\tau_{N{-}3}$ out-competes sequential tunneling.  (c) Fits to zero-bias gate traces yield $\lambda_{q}$ and $\bar{\Gamma}$ values, here $\lambda_{q} = 2.7$, $\bar{\Gamma}=$\SI{40}{\micro\electronvolt}. Traces are offset for clarity. (d) Modelled conductance map at 5 K, using $\hbar\omega_{q} = 9$~meV, $\lambda_{q} = 2.7$, $\bar{\Gamma}=$\SI{40}{\micro\electronvolt}, $1/\tau_{N-4} = 100 \times W^{N-4,N-3}_{1,0}$, $1/\tau_{N-3} = 1/100 \times W^{N-3,N-4}_{1,0}$; sequential tunneling rates are calculated at the red and grey points respectively.
\label{theory}}
\end{figure}

As shown in Fig. \ref{temp_depen}b, the sidebands do not cross the zero-bias axis. At 5 K the conductance of the first sideband (Fig. \ref{temp_depen}b, grey) is suppressed below $|eV_{\rm{sd}}| = \hbar\omega_{q}$, and the second sideband conductance (pink) is suppressed below $2\hbar\omega$. This suggests that the features within the \nfour~ Coulomb diamond are due to cotunneling-assisted sequential tunneling processes\citep{gaudenzi17, Lueffe2008}. The emergence of the first sideband is shown in Fig. \ref{theory}a. At low voltages ($|eV_{\rm{sd}}| < \hbar\omega_{q}$) cotunneling must be elastic ($q'=q$) and will contribute a small portion of tunneling current (cotunneling is second-order in $\Gamma$).  Above $|eV_{\rm{sd}}| = \hbar\omega_{q}$ inelastic cotunneling events, which leave the molecule in an excited vibrational state, are energetically allowed (Fig. \ref{theory}a). Under weak molecule-electrode coupling, inelastic cotunneling processes carry little net current, and we do not observe gate-independent co-tunneling lines within the Coulomb diamonds\citep{franceschi01}. If the vibrational excitation that results from inelastic cotunneling is slow to relax compared to the sequential tunneling rates, $W^{N,N'}_{q,q'}$ (where $N'= N\pm1$), specifically $W^{N{-}4,N{-}3}_{1,0}$, then a sequential tunneling pathway to the ground state of \nthree~is opened up (Fig. \ref{theory}a). The $\ket{N{-}3,q={0}}$ state can subsequently undergo sequential tunneling to up to the second vibrationally excited state of \nfour. If electron-phonon coupling, parameterized by the coupling constant: $\lambda_{q}$, is weak ($\lambda_{q} < 1$) then the transition to the $\ket{N{-}4,q{=}0}$ is most likely, and the molecule returns to the vibrational ground state until another inelastic cotunneling begins the cycle again. If $\lambda_{q} > 1$ then transitions to $\ket{N{-}4,q{=}1,2}$ out-compete the return to the \nfour~ground state, and a single inelastic cotunneling event leads to sustained sequential electron tunneling through the molecule, even in the CB region, via vibrationally excited states of \nfour. Therefore the prominence of the sidebands is enhanced when electron-phonon coupling is strong.

We look to understand these mechanisms quantitatively. Conductance is a sum of the cotunneling and sequential tunneling contributions, i.e.,  $G_{sd} \approx G_{seq} + G_{cot}$. At low temperature and zero-bias these are given analytically by\citep{Koch2006}:

\begin{equation}
    G_{seq} = -\frac{2e^2}{\hbar}\frac{\Gamma_{L}\Gamma_{R}}{\Gamma_{L}+\Gamma_{R}}\frac{f'(\epsilon_{d})}{1+f(\epsilon_{d})}\rm{e}^{-\lambda^2}
\end{equation}

and

\begin{equation}
    G_{cot} = \frac{2e^2}{h}\frac{\Gamma_{L}\Gamma_{R}}{1+f(\epsilon_{d})}\left[\frac{1-f(\epsilon_{d})}{(\epsilon_{d} + \lambda^{2}\hbar\omega)^2}+\frac{2f(\epsilon_{d})}{(\epsilon_{d}-\lambda^2\hbar\omega)^2}\right]
\end{equation}

where $\Gamma_{L}$ and $\Gamma_{R}$ are the couplings of the molecule to left and right electrodes, $f(\epsilon_{d})$ is the Fermi distribution at the energy of the molecular transition ($\epsilon_{d}$), and $f'(\epsilon_{d})$ is its derivative. We fit zero-bias conductance traces (at 10 K and below) to $\bar{\Gamma}$ and $\lambda_{q}$, shown in Fig. \ref{theory}c ($\bar{\Gamma} = \Gamma_{L}\Gamma_{R}/(\Gamma_{L} + \Gamma_{R})$). The two parameters cannot be determined independently (they are positively covariant\footnotemark[\value{footnote}]), although it is possible to determine the range in which they lie. Sequential tunneling is exponentially suppressed in $\lambda_{q}$, however cotunneling can still occur via highly excited virtual states of \nthree, therefore at large $\lambda_{q}$ (> 2.8), and with the corresponding increase in $\bar{\Gamma}$, there are significant contributions from elastic cotunneling that are not observed experimentally. This sets an upper bound for $\lambda_{q}$. Simulations (\textit{vide infra}) of the conductance map show at $\lambda_{q} < 2$ (corresponding to $\bar{\Gamma} <$\SI{1}{\micro\electronvolt}) the sidebands become too low in conductance to observe. The lack of sidebands at low values of $\lambda_{q}$ and $\bar{\Gamma}$ results from reduced inelastic cotunneling rates that initiate the sidebands, along with the enhanced rates of sequential transfer from excited vibrational states of $\ket{N{-}3,q}$ to the ground state, $\ket{N{-}4,q={0}}$ due to larger Franck-Condon factors. Therefore, we find pairs of parameters within ranges of $2-3$ for $\lambda_{q}$ and 1--\SI{100}{\micro\electronvolt} for $\bar{\Gamma}$. Theoretical work predicted that slow vibrational relaxation coupled with values $\lambda_{q}$ in the range of 2--3 would result in negative differential conductance (NDC) after the conductance peaks of the vibrational sidebands, due to the depopulation of non-equilibrium vibrational distribution to the ground state by sequential transfer as the sidebands enter the bias window\citep{Leijnse2008}. Our experimental observation of NDC (Fig. \ref{temp_depen}a) and calculated range of $\lambda_{q}$ are consistent with this prediction.
The criteria for the emergence of the first \nfour~sideband is: $1/\tau_{N{-}4} \leq W^{N{-}4,N{-}3}_{1,0}$. Therefore a calculation of the sequential rate yields a lower bound for the vibrational relaxation time, $\tau_{N{-}4}$,  within the junction. Sequential tunneling rates are given by\citep{Koch2006}:

\begin{equation}
    W^{N-4,N-3}_{q,q';a} = 2\Gamma_{a}|M_{q,q'}|^2f_{a}(\epsilon + [q' - q]\hbar\omega)
\label{eqn:wcot}
\end{equation} 

where $M_{q,q'}$ are the Franck-Condon matrix elements. Across the range of $\lambda_{q}, \bar{\Gamma}$, the calculated values of $W^{N{-}4,N{-}3}_{1,0;L}$ are in the narrow range of 56 -- 120 MHz,\footnotemark[\value{footnote}] giving a lower bound for the vibrational relaxation time of $\tau_{N{-}4} > 8$ ns. The absence of sidebands within the \nthree~charge state is an intriguing part of the system dynamics. If $1/\tau_{N{-}3} > W^{N{-}3,N{-}4}_{1,0}$ the sidebands are not present (Fig. \ref{theory}b) pointing to a charge-state dependence of the vibrational relaxation rates. Using a minor extension to the model developed by Koch \textit{et al.} to include charge-state dependent relaxation times, $\tau$, we calculate the conductance map (Fig. \ref{theory}d) using these inequalities\citep{Koch2006}, which shows good correspondence with the experimental data, reproducing the key features of the transport spectrum.

The temperature dependence of the conductance supports the assignment of the sidebands as cotunneling-assisted sidebands. At zero bias the Coulomb peak and the first sideband follow equilibrium behavior (Fig. \ref{temp_depen}c). The conductance of the Coulomb peak ($V_{\rm{CP}} = -1.08$~V) decays as $G_{max} \propto \frac{1}{k_{B}T}$, due to thermal broadening of the electrode Fermi-Dirac distributions, and the conductance at the point where the first sideband would cross $V_{sd} = 0$ (from which electron transfer is driven from $q=1$, red circle, Fig. \ref{temp_depen}a) can be fitted to the product of thermal broadening and the Bose distribution\citep{Leturcq2009}:

\begin{equation}
\label{eqn:bose}
  G_{max} \propto \frac{1}{k_{B}T} \times  \frac{1}{exp(\hbar\omega_{q}/k_{B}T)-1}
\end{equation}

with $\hbar\omega_{q} = 8 \pm 1$~meV. This demonstrates the zero-bias conductance increase is driven thermally by an increasing equilibrium population in $q=1$. The picture is different at higher bias however (Fig. \ref{temp_depen}d). At $eV_{\rm{sd}} = \hbar\omega_{q}$ the conductance of the first sideband is already high at 5 K and decreases with increasing temperature, indicating there is a substantial non-equilibrium population at $eV_{\rm{sd}} = \hbar\omega_{q}$, that must be driven by the tunneling dynamics of the system, i.e. inelastic co-tunneling and slow relaxation. For the second sideband (orange line in Fig. \ref{temp_depen}) sequential tunneling begins from $\ket{N{-}4,q{=}2}$. At $|eV_{\rm{sd}}| = \hbar\omega_{q}$ (red square), this state can only arise as the result of two inelastic excitations of the vibrational ground state or due to an inelastic excitation of a thermally-populated $\ket{N-4,q{=}1}$ state. The latter becomes appreciable only at higher temperature. Since the emergence of the second sideband at positive bias is observed only above 10 K, this suggests the second mechanism dominates. A fit to equation \ref{eqn:bose} gives $\hbar\omega_{q} = 7.4 \pm 1.5$~meV. The sidebands are stronger at negative bias, likely due to asymmetries resulting from both molecule-electrode coupling, common in single-molecule devices\cite{limburg18}, and an asymmetric voltage drop across the junction ($\alpha_{\rm{S}} = 0.24$, extracted from the Coulomb diamonds slopes).

\begin{figure}
\includegraphics{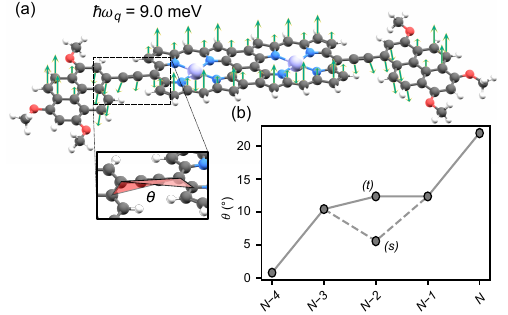}
\caption{(a) An example of vibrational mode around the energy (9~meV), the energy responsible for FC blockade in \nthree/\ntwo~ and the sidebands in \nfour/\nthree. (b) DFT-calculated dihedral angle, $\theta$ (shown in (a)), as a function of charge state. For \ntwo, the closely spaced singlet and triplet (the ground state) are labelled by \textit{(s)} and \textit{(t)} respectively. 
\label{dft}}
\end{figure}

The neighboring transition, \nthree/\ntwo~(blue box on Fig. \ref{schematics}d), as with \nfour/\nthree~, displays signatures of strong electron-phonon coupling, i.e., Franck-Condon blockade involving a mode of 10 meV\footnotemark[\value{footnote}]. Vibrational modes that couple strongly to the electron transfer process are those that displace atoms along the same vectors that define the nuclear rearrangement. Density functional theory calculations aid our understanding of strong electron-vibrational coupling in the oxidations of \textbf{FP2}\citep{g16}. Frequency calculations (B3LYP/6-31G(d) functional/basis set) on \textbf{FP2} optimized in the $N$ to \nfour~ states show that modes around 9 meV correspond to out-of-plane motions of the molecule, i.e., saddling of the porphyrin unit, or torsional motions between the pyrene/porphyrin $\pi$-systems. Fig. \ref{dft} shows the angle between the $\pi$-systems indeed vary with charge state, providing an explanation for why a mode of this nature would couple strongly to charge transfer in the oxidations of \textbf{FP2}, and similar motions were observed to couple to electron transfer in porphyrin monomers with pyrene anchoring groups\citep{Thomas2019}.

The origins of slow vibrational relaxation are more difficult to pin down; dissipation is affected by energy-dependent coupling to the phononic background of the substrate and electrodes, as well as intramolecular vibrational redistribution\citep{franke12, nitzan73}. A room-temperature study of molecular vibrational lifetimes in anthracene derivatives in molecular junctions gave values in the 10--100 picosecond range, similar to those measured in solution, where picosecond relaxation is typical\citep{laubereau72}. In general vibration relaxation will slow as temperature decreases\citep{nitzan73}. For example, our measured relaxation time of $\tau_{N{-}4} > 8$ ns, giving a $Q$-factor ($\tau E / h$) of $>20,000$, is comparable to that found for the radial breathing mode of a carbon nanotube at 5 K\citep{leroy04}. Particularly interesting is the link between the molecular charge state, and consequently, its geometry, and the vibrational relaxation rate. It is noteworthy that \textbf{FP2} in the \nfour~ oxidation state has a coplanar porphyrin-pyrene system (Fig. \ref{dft}). Therefore the potential resulting from rotation around the torsional angle, $\theta$, is approximately harmonic (an assumption on which the above analysis is based), whereas for $N$ to \nthree~ with non-zero values of $\theta$ the potential is quartic with two local minima, corresponding to \textit{syn} and \textit{anti} conformations of the pyrene-porphyrin-pyrene system. Quartic potentials increase the energy density of low-energy modes and, combined with the asymmetry induced by the substrate, could expedite intramolecular vibrational relaxation for the $N$ to \nthree~ states\citep{bethardy94,sun21}.

In summary, we demonstrate that low-energy, torsional motions of a pyrene-porphyrin dimer-pyrene couple strongly to electron transfer and can lead to cotunneling-assisted absorption sidebands. We observe a strong dependence of the vibrational relaxation rate on the charge state of the molecule, and moving beyond the harmonic approximation may be required to fully unravel the tunneling dynamics at play. This study of non-equilibrium vibrational dynamics and dissipation sheds light on heat and energy flow in molecular devices.

\begin{acknowledgments}
This work was supported by the EPSRC (grants EP/N017188/1 and EP/R029229/1). JAM acknowledges funding from the Royal Academy of Engineering and a UKRI Future Leaders Fellowship, Grant No. MR/S032541/1. The authors would like to acknowledge the use of the University of Oxford Advanced Research Computing (ARC) facility in carrying out this work. http://dx.doi.org/10.5281/zenodo.22558
\end{acknowledgments}

\bibliographystyle{aipnum4-1}
\bibliography{FusedDimer}
\clearpage
\pagebreak
\onecolumngrid

\section{Supplementary Information}

\renewcommand{\thefigure}{S\arabic{figure}}
\setcounter{figure}{0}
\subsection{Synthetic details}

\textbf{FP2}. The dibromo- precursor Br\textsubscript{2}FP2 (6.88 mg, 1.99 µmol) and 1-ethynyl-3,6,8-tridodecoxypyrene (7.79 mg, 10 µmol) were dissolved in toluene (4 mL) and DIPA (1 mL). The solution was degassed by freeze-pump-thaw three times before [Pd(PPh\textsubscript{3})\textsubscript{4}] (0.23 mg, 0.2 µmol) and CuI (0.04 mg, 0.2 µmol) were added under a flow of argon. After another freeze-pump-thaw cycle, the mixture was heated to 50 \textdegree C for 20 hours. The solvents were removed and the product purified by repeated chromatography (SiO\textsubscript{2}, PE/DCM 5:1 followed by size-exclusion, CHCl\textsubscript{3}), to give \textbf{FP2} as a green/brown solid (\SI{6.38}{\milli\gram}, yield = 64\%). For deposition onto graphene electrodes \SI{2}{\micro\litre} of a \SI{2}{\micro\molar} toluene solution of \textbf{FP2} was dropcast onto the devices.\
\textsuperscript{1}H NMR (400 MHz, CDCl\textsubscript{3}) $\delta$ 8.67 (d, J = 4.6 Hz, 4H), 8.50 (d, J = 9.3 Hz, 2H), 8.31 (d, J = 9.3 Hz, 2H), 8.26 (d, J = 9.4 Hz, 2H), 8.18 (d, J = 9.3 Hz, 2H), 7.78 (s, 8H), 7.76 (s, 4H), 7.75 (s, 2H), 7.49 (d, J = 4.5 Hz, 4H), 7.06 – 7.05 (m, 6H), 4.32 – 4.19 (m, 12H), 2.02 – 1.86 (m, 12H), 1.63 – 1.50 (m, 12H), 1.43 – 1.01 (m, 288H), 0.86 – 0.64 (m, 138H).\
m/z (MALDI-TOF, dithranol): 4843.6958 ([M]+ calcd. 4860.95).

\subsection{Additional transport data}

\begin{figure}[!htb]
\includegraphics[width=0.8 \textwidth]{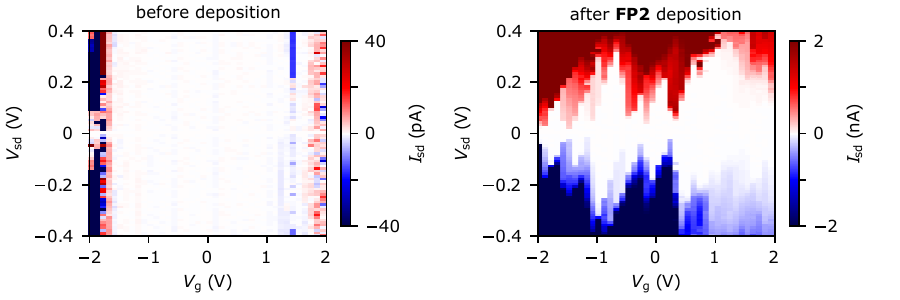}
\caption{Current stability diagrams, measured at 4 K, before and after \textbf{FP2} deposition.
\label{beforeafter}}
\end{figure}

\clearpage

\subsection{Determination of ranges of molecule-electrode and electron-phonon coupling}

\begin{figure}[!htb]
\includegraphics[width=0.7 \textwidth]{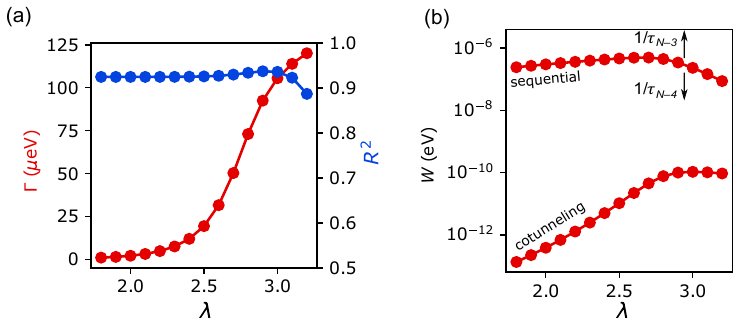}
\caption{(a) $R^{2}$ (blue markers) of fit for zero-bias gate traces from 5 K to 10 K, with different values of $\lambda_{q}$, the corresponding molecule-electrode coupling is given in red. (b) For each set of $\lambda_{q}$ and $\bar{\Gamma}$ the calculated sequential tunneling rate $W^{N-4,N-3}_{1,0;L}$ and cotunneling rates $W^{N-4,N-4}_{0,1}$ at the onset of the first sideband in \nfour.
\label{fittingInfo}}
\end{figure}

\subsection{Franck-Condon blockade in \nthree/\ntwo}

\begin{figure*}[!htb]
\includegraphics[width=0.6 \textwidth]{SI_FCblockade.pdf}
\caption{(a) Conductance maps at different temperatures of the \nthree/\ntwo~ transition (blue box in Fig. \ref{schematics}), showing conductance suppression at low bias and low temperature. (b) Zero-bias gate traces displaying the lifting of the blockade with increasing temperature, consistent with FC blockade, and a fit the temperature-dependence of the Coulomb peak (equation \ref{eqn:bose}) gives the mode energy of $\lambda_{q}$.
\label{FC_SI}}
\end{figure*}

\end{document}